\documentclass[%
 superscriptaddress,
 twocolumn,
 nofootinbib,
 amsmath,amssymb,
 aps,
 prd,
]{revtex4-1}

\usepackage[colorlinks]{hyperref}
\usepackage{tabularx}
\usepackage{graphicx}
\usepackage{amssymb,amsmath,bm,tensor,braket}
\usepackage{xcolor}
\usepackage[varg]{txfonts}
\usepackage{enumerate}
\usepackage{mathtools}
\usepackage[capitalize]{cleveref}
\usepackage[utf8]{inputenc}
\usepackage[normalem]{ulem}
\usepackage{esvect}
\usepackage{dcolumn}

\usepackage{subfigure}
\usepackage[normalem]{ulem}

\newcommand{\Reply}[1]{#1}

\begin{document}

\preprint{APS/123-QED}

\title{Moment of Inertia for Axisymmetric Neutron Stars in the Standard-Model Extension}

\author{Yiming Dong}
\affiliation{Department of Astronomy, School of Physics, Peking University, Beijing 100871, China}%
\affiliation{Kavli Institute for Astronomy and Astrophysics, Peking University, Beijing 100871, China}

\author{Zexin Hu}%
\affiliation{Department of Astronomy, School of Physics, Peking University, Beijing 100871, China}%
\affiliation{Kavli Institute for Astronomy and Astrophysics, Peking University, Beijing 100871, China}

\author{Rui Xu}
\affiliation{Department of Astronomy, Tsinghua University, Beijing 100084, China}%
\affiliation{Kavli Institute for Astronomy and Astrophysics, Peking University, Beijing 100871, China}%

\author{Lijing Shao}\email[Corresponding author: ]{lshao@pku.edu.cn}
\affiliation{Kavli Institute for Astronomy and Astrophysics, Peking University, Beijing 100871, China}%
\affiliation{National Astronomical Observatories, Chinese Academy of Sciences, Beijing 100012, China}

\date{\today}

\begin{abstract}
We develop a consistent approach to calculate the moment of inertia (MOI) for
axisymmetric neutron stars (NSs) in the Lorentz-violating Standard-Model
Extension (SME) framework. To our knowledge, this is the first relativistic MOI
calculation for axisymmetric NSs in a Lorentz-violating gravity theory other
than deformed, rotating NSs in the General Relativity. Under Lorentz
violation, there is a specific direction in the spacetime and NSs get stretched
or compressed along that direction. When a NS is spinning stationarily along
this direction, a conserved angular momentum and the concept of MOI are well
defined. In the SME framework, we calculate the partial differential equation
governing the rotation and solve it numerically with the finite element method
to get the MOI for axisymmetric NSs caused by Lorentz violation. Besides, we
study an approximate case where the correction to the MOI is regarded solely
from the deformation of the NS and compare it with its counterpart in the
Newtonian gravity. Our formalism and the numerical method can be extended to
other theories of gravity for static axisymmetric NSs.
\end{abstract}


\maketitle



\section{Introduction}
\label{Sec1_Introduction}

At the classical level, gravitational phenomena are well described by the
General Relativity (GR), which has withstood various experimental tests over the
past century with flying colors~\cite{Will:2014kxa, Will:2018bme}. At the
quantum level, the Standard Model (SM) of particle physics provides an accurate
description of interactions between microscopic particles. Together, GR and SM
form the foundation for our contemporary understanding of the nature. However,
there has been a longstanding quest to a final theory, the so-called quantum
gravity, that can consistently describe all phenomena. Quantum gravity is
expected to exhibit unique behaviors different from GR at the Planck energy
scale, but testing  theories at the Planck  scale is challenging if
possible~\cite{Gambini:1998it, Amelino-Camelia:2008aez}. Therefore, physicists
have turned attention to searching for relic effects of quantum gravity at low
energy scales, and Lorentz violation is one possible relic
effect~\cite{Kostelecky:1988zi, Kostelecky:1989jw, Gambini:1998it,
Kostelecky:2002hh, Bailey:2006fd, Amelino-Camelia:2008aez}. A field-theoretic
approach, the Standard-Model Extension (SME), collects all possible operators of
Lorentz violation in a Lagrangian~\cite{Colladay:1996iz, Colladay:1998fq,
Kostelecky:2003fs, Bailey:2006fd},
\begin{equation}
    \mathcal{L}_{\rm SME} = \mathcal{L}_{\rm GR} + \mathcal{L}_{\rm SM} +
    \mathcal{L}_{\rm LV} + \mathcal{L}_{\rm k}\,,
\end{equation}
where $\mathcal{L}_{\rm GR}$ represents the Einstein-Hilbert term for GR,
$\mathcal{L}_{\rm SM}$ is the Lagrangian of the SM, $\mathcal{L}_{\rm LV}$ is
the Lorentz-violating term, and $\mathcal{L}_{\rm k}$ describes the dynamics of
the Lorentz-violating fields. For the term $\mathcal{L}_{\rm LV}$, in this study
we consider the minimal gravitational Lorentz violation with operators of
mass-dimension four~\cite{Bailey:2006fd},
\begin{equation}
    \mathcal{L}^{(4)}_{\rm LV} = \frac{1}{16 \pi} \left(-u R + s^{\mu\nu} R^{\rm
    T}_{\mu\nu} + t^{\alpha\beta\gamma\delta}C_{\alpha\beta\gamma\delta}
    \right)\,,
\end{equation}
where $R$ is the Ricci scalar, $R^{\rm T}_{\mu\nu}$ is the trace-free Ricci
tensor, $C_{\alpha\beta\gamma\delta}$ is the Weyl conformal tensor, and $u$,
$s^{\mu\nu}$, $t^{\alpha\beta\gamma\delta}$ are the Lorentz-violating fields. In
the SME framework, we can describe the Lorentz-violating fields by introducing
their vacuum expectation values, $\bar{u}$, $\bar{s}^{\mu\nu}$, and
$\bar{t}^{\alpha\beta\gamma\delta}$, which are then called the Lorentz violation
coefficients~\cite{Bailey:2006fd}. Extensive experiments have been conducted to
constrain the Lorentz violation coefficients~\cite{Chung:2009rm, Shao:2014oha,
Shao:2014bfa, Yunes:2016jcc, Bourgoin:2017fpo, Shao:2018vul, Shao:2018lsx,
Shao:2019cyt, Shao:2020shv, Kostelecky:2008ts}.

In this work we will consider neutron stars (NSs) in the SME framework. NSs are
ideal laboratories for testing fundamental theories and principles, including
the Lorentz symmetry~\cite{Taylor:1979zz, Kramer:2006nb, Shao:2016ezh,
Kramer:2016kwa, Miao:2020wph, Kramer:2021jcw, Shao:2022izp, Hu:2023vsq}.
Pulsars, which are rotating NSs, provide us a superb opportunity to test
theories of gravity~\cite{Taylor:1979zz, Kramer:2006nb, Kramer:2021jcw}
including the Lorentz symmetry in circumstances of strong gravitational
field~\cite{Shao:2014oha, Shao:2014bfa, Shao:2018lsx, Shao:2019cyt}.  In some
cases, the uncertain equation of state (EOS) for dense nuclear matter of NSs
could introduce degeneracy with gravity tests~\cite{Shao:2017gwu, Shao:2019gjj,
Shao:2022koz}.  Nevertheless, measurements of NS properties, such as mass,
radius, moment of inertia (MOI), and tidal Love number offer us an avenue to
study the EOS~\cite{Akmal:1998cf, Lattimer:2000nx, Lattimer:2006xb,
Hinderer:2009ca, Demorest:2010bx, Ozel:2016oaf, NANOGrav:2019jur, Li:2020wbw}.
Through high-precision pulsar timing observations~\cite{Demorest:2010bx,
NANOGrav:2019jur, Hu:2020ubl, Kramer:2021jcw}, gravitational-wave detections of
binary NS mergers~\cite{LIGOScientific:2017vwq, LIGOScientific:2018cki,
De:2018uhw}, and multi-wavelength observations of X-ray
pulsars~\cite{Riley:2021pdl, Raaijmakers:2021uju}, we can obtain high-precision
measurements of the structure of NSs. These constraints on the EOS of NSs help
us gain insights into the physics of dense nuclear matter, as well as gravity
tests.

With Lorentz violation, NSs undergo non-spherical deformations. Studying the
structure of NSs under Lorentz violation can help us test the SME framework and
offer the potential for identifying additional observable
effects~\cite{Xu:2019gua, Xu:2020zxs, Xu:2021dcw}. \citet{Xu:2019gua} employed a
method similar to the post-Tolman-Oppenheimer-Volkoff (post-TOV)
approach~\cite{Glampedakis:2015sua} to deal with the effects from Lorentz
violation, and obtained the leading-order corrections to the structure of NSs
caused by Lorentz violation. \Reply{In this paper, we attempt to extend the study of the structure of NSs under
Lorentz violation. In particular, we focus on the MOI of NSs.}

\Reply{MOI is one of the crucial
structural parameters of NSs, as it characterizes the rotational properties of NSs. MOI is closely connected to the central issues in NS physics. Firstly, Observations and studies of MOI can help us constrain the EOS of NSs~\cite{Bejger:2002ty, Morrison:2004df, Kramer:2021jcw}. MOI of NSs varies with different EOSs, and it can also be measured directly from high-precision observations of binary pulsars~\cite{Hu:2020ubl, Kramer:2021jcw}. Owing to the high precision of pulsar timing, there is the potential to detect orbital effects related to the MOI, e.g.\ through the periastron advance caused by the spin-orbit coupling. Currently, with a 16-year data span, an upper limit of the MOI for PSR~J0737$-$3039A in the Double Pulsar system has been obtained~\cite{Kramer:2021jcw}. With the advent of the next generation radio telescopes, such as the Square Kilometre Array (SKA), there is hope for direct measurements of the MOI of NSs~\cite{Smits:2008cf, Hu:2020ubl}, offering us a means to study the EOS of NSs. Secondly, glitch phenomena in pulsar timing observations are also believed to be
related to the MOI of NSs. Glitches are one type of timing irregularities in
pulsar timing observations, which manifest as sudden changes in rotation
frequencies of pulsars, and are often followed by a
relaxation~\cite{Anderson:1975zze, Espinoza:2011pq, Yu:2012mp}. There are a lot
of theoretical models that aim to explain glitches, such as models associated
with superfluid and crustquake~\cite{Andersson:2002zd, Haskell:2015jra,
Lai:2023axr, Yim:2023nda}. Investigating the origins of glitches contributes to our
comprehension of the physics within NSs. Considering that the angular momentum of a NS is conserved or almost conserved, any changes in the MOI will result in variations in the angular velocity, leading to noticeable observational effects for pulsars. If we intend to explain glitches
through deformations of NSs, we need to calculate the MOI and infer the
variation in the MOI from the change of angular velocity. Some studies have also
attempted to explain the unexplained issues in glitches, such as the deficiency
of MOI contributed by the NS crust~\cite{Andersson:2012iu}, e.g.\ with a
modified gravity~\cite{Staykov:2015mma}. Additionally, considering the precision of pulsar spin measurements, other MOI-related physical processes affecting NS rotation may also be measurable. In this context, calculating the MOI corrections induced by these physical processes is essential.}

Research on the MOI of NSs in a relativistic setting can be traced back to 1960s
when Hartle and Thorne~\cite{Hartle:1967he, Hartle:1968si} calculated the
structure of slowly rotating NSs in GR and computed the MOI for spherically
symmetric NSs. Their results showed a significant difference between the
calculations in GR and those in the Newtonian gravity. Another important
theoretical work related to the MOI of NSs is the discovery of the so-called
I-Love-Q relation, which is one of the most famous universal relations for
NSs~\cite{Yagi:2013bca, Yagi:2013awa}. Numerical calculations revealed that the
relations between any two of the dimensionless MOI, the dimensionless tidal Love
number, and the dimensionless quadrupole moment are insensitive to the EOS of
NSs. The I-Love-Q relation provides us a way to test gravity theories
independently of the EOS~\cite{Shao:2022koz}. In addition, calculations have
also been performed on the MOI for NSs in alternative gravity
theories~\cite{Staykov:2014mwa, Pani:2014jra, Yazadjiev:2016pcb}, but they are
limited to the assumption of the spherical background configuration.

It is worth noting that previous calculations of the MOI of NSs have based on
the assumption of spherical symmetry. To our knowledge, no calculations in the
relativistic setting have been performed yet regarding the correction to the MOI
caused by non-spherical deformations other than rotation itself. Indeed,
considering the MOI of non-spherical NSs is meaningful. Firstly, there exist
various physical processes that can induce non-spherical deformations in NSs,
such as crustal deformations, magnetic field effects and so
on~\cite{Haskell:2007bh, Lander:2014csa}. Exploring the corrections to MOI
caused by non-spherical deformations can provide valuable insights into the
structure and dynamics of NSs, and contribute to our understanding of complex
behaviors of NSs. Secondly, from the perspective of gravity theories, there are
some modified gravity theories breaking the spherical symmetry, such as the
bumblebee theory~\cite{Bailey:2006fd} and the Einstein-\AE{}ther
theory~\cite{Jacobson:2000xp}. In these gravity theories, there may exist
axisymmetric solutions for NSs which are more stable than the spherical ones. In
that case, studying the structure of non-spherical NSs helps us understand these
theories better.

In this context, we present a consistent calculation of the MOI for axisymmetric
NSs in the SME framework. The organization of the paper is as follows. In
Sec.~\ref{Sec2_MOI_GR}, we introduce the calculation of MOI for spherical NSs in
GR to lay the groundwork. In Sec.~\ref{Sec3_MOI_SME}, we first review the
deformed NSs in the SME found in Ref.~\cite{Xu:2019gua} in
Sec.~\ref{Sec31_NS_SME}. Then in Sec.~\ref{Sec32_Mod_PDE}, we obtain the partial
differential equation (PDE) that describes the rotational metric in the SME,
retaining the correction terms up to the first order in the Lorentz violation
coefficients. In Sec.~\ref{Sec33_Num_Cal}, we solve the PDE numerically with the
finite element method to get the MOI for NSs. Finally, we summarize in
Sec.~\ref{Sec5_Summary}. In this paper, we adopt the units where $G=c=1$.


\section{MOI of spherical NSs in GR}
\label{Sec2_MOI_GR}

In GR, the definition of MOI is based on the definitions of angular velocity and
angular momentum. To obtain the MOI of a NS, we need to calculate the
gravitational field equation to get the metric of the rotating
spacetime~\cite{Hartle:1967he, Hartle:1968si}. We begin with the metric of a
stationary, axially symmetric system,
\begin{equation}
    d s^2=-H^2 d t^2+Q^2 d r^2+r^2 K^2\left[d \theta^2+\sin ^2 \theta(d
    \varphi-L d t)^2\right]\,,
\label{Eq_metric_Ax}
\end{equation}
where $H$, $Q$, $K$, and $L$ are functions of $r$ and $\theta$. The
corresponding four-velocity of the fluid reads,
\begin{equation}
    u^\mu=(u^t,0,0, \Omega u^t)\,.
\label{Eq_FourV}
\end{equation}

We adopt the assumption of slow rotation, where the effects on pressure, energy
density, and gravitational field caused by the rotation can be treated as
perturbations. In this case, we can expand $L$ in orders of $\Omega$,
\begin{equation}
    L(r, \theta)=\omega(r, \theta)+\mathcal{O}\left(\Omega^3\right)\,.
\label{Eq_L_exp}
\end{equation}
Then, we can solve $\omega$ from the following field equation,
\begin{equation}
    G^{t}_{\varphi}=8\pi T^{t}_{\varphi}\,.
\label{Eq_field}
\end{equation}
It is worth noting that the leading-order correction to $L$ is of order $\Omega$
but the leading-order corrections of $H$, $Q$, and $K$ are of order $\Omega^2$.
If we want to calculate the leading-order effect, we can consider all diagonal
components of the metric in Eq.~(\ref{Eq_field}) as the background solution of a
spherical NS, whose metric is commonly written as
\begin{equation}
    d s^2=-e^{\nu(r)} d t^2+e^{\lambda(r)} d r^2+r^2\left(d \theta^2+\sin ^2
    \theta d \varphi^2\right)\,.
\label{Eq_metric_TOV}
\end{equation}
Finally, we can express the field equation~(\ref{Eq_field}) in the form
of~\cite{Hartle:1967he, Hartle:1968si}
\begin{equation}
    \frac{1}{r^4}\frac{\partial}{\partial r}\left(r^4 j \frac{\partial
    \bar{\omega}}{\partial r} \right) + \frac{4}{r} \frac{\mathrm{d}
    j}{\mathrm{d} r} \bar{\omega} +
    \frac{e^{(\lambda-\nu)/2}}{r^2}\frac{1}{\sin^{3}{\theta}}
    \frac{\partial}{\partial \theta} \left( \sin^{3}{\theta} \frac{\partial
    \bar{\omega}}{\partial \theta} \right)=0\,,
\label{Eq_GR_MOI}
\end{equation}
where
\begin{equation}
    \bar{\omega}(r, \theta) \equiv \Omega-\omega(r, \theta)\,,
\label{Eq_defination_baromega}
\end{equation}
and $j(r)\equiv e^{(\lambda+\nu)/2}$. The regular condition and boundary
condition are given by
\begin{align}
    \bar{\omega}\big|_{r=0} &= {\rm constant}\,,
\label{Eq_Bound1} \\
    r^{2}\left(\Omega-\bar{\omega}\right)\big|_{r\rightarrow\infty} &= 0\,.
    \label{Eq_Bound2}
\end{align}

The separation of variables method is the most straightforward approach to solve
the PDE. Fortunately, Eq.~(\ref{Eq_GR_MOI}) can be separated using vector
spherical harmonics,
\begin{equation}
    \bar{\omega}(r,\theta) = \sum_{l=1}^{\infty} \bar{\omega}_{l}(r)
    \left(-\frac{1}{\sin{\theta}} \frac{\mathrm{d}
    P_{l}(\cos{\theta})}{\mathrm{d} \theta} \right)\,,
\label{Eq_Vec_har}
\end{equation}
where $P_{l}(\cos{\theta})$ is the Legendre function, and $\bar{\omega}_{l}(r)$
satisfies,
\begin{equation}
    \frac{1}{r^{4}} \frac{\mathrm{d}}{\mathrm{d} r}
    \left[r^{4}j(r)\frac{\mathrm{d} \bar{\omega}_{l}}{\mathrm{d} r}\right] +
    \left[\frac{4}{r} \frac{\mathrm{d} j}{\mathrm{d} r} - e^{(\lambda-\nu)/2}
    \frac{l(l+1)-2}{r^2} \right] \bar{\omega}_{l} = 0\,.
\label{Eq_MOI_GR_Sep}
\end{equation}
With the boundary condition and the regular condition, it is proved that
$\bar{\omega}_{l}(r)$ vanishes except for $l=1$ and thus $\bar{\omega}$ is
independent of $\theta$~\cite{Hartle:1967he,Hartle:1968si}.
Equation~(\ref{Eq_GR_MOI}) reduces to an ordinary differential equation (ODE),
\begin{equation}
    \frac{1}{r^4} \frac{d}{d r}\left(r^4 j \frac{d \bar{\omega}}{d
    r}\right)+\frac{4}{r} \frac{d j}{d r} \bar{\omega}=0\,.
\label{Eq_ODE_MOI_GR}
\end{equation}
The solution outside the star has the form
\begin{equation}
    \bar{\omega}(r)=\Omega-\frac{2 J}{r^3}\,,
\label{Eq_baromega_Solution}
\end{equation}
where $J$ is just the angular momentum of the NS. Finally, the definition of MOI
reads $I \equiv {J}/{\Omega}$.

We can get a more compact equation for the MOI with some further definitions.
First, we define an angular momentum function of the variable $r$, 
\begin{equation}
     \mathcal{J}(r) \equiv \frac{1}{6} r^4\left(\frac{d \bar{\omega}(r)}{d
     r}\right)\,,
\end{equation}
and a corresponding MOI function
\begin{equation}
    \mathcal{I}(r) \equiv \frac{\mathcal{J}(r)}{\Omega}\,.
\label{Eq_Solution_GR}
\end{equation}
When $r$ is larger than the radius of the NS, $R$, the angular momentum function
$\mathcal{J}$ and the MOI function $\mathcal{I}$ equal to $J$ and $I$,
respectively.

With the above definitions and Eq.~(\ref{Eq_ODE_MOI_GR}), we can obtain the ODE
of $\mathcal{I}(r)$~\cite{Hu:2023vsq},
\begin{equation}
    \frac{\mathrm{d} \mathcal{I}}{\mathrm{d} r}=\frac{8}{3} \pi r^4
    \rho\left(1+\frac{p}{\rho}\right)\left(1-\frac{5}{2}
    \frac{\mathcal{I}}{r^3}+\frac{\mathcal{I}^2}{r^6}\right)\left(1-\frac{2
    m}{r}\right)^{-1}\,,
\label{Eq_dIdr}
\end{equation}
where $\rho$ and $p$ are the energy density and the pressure of NSs
respectively, $m$ is the mass function defined in the TOV equation. The form of
Eq.~(\ref{Eq_dIdr}) is similar to the TOV equation for $p$ where the right-hand
side is the Newtonian term times three dimensionless factors.

\begin{figure}
    \centering
    \includegraphics[scale=0.43, trim=10 30 0 50]{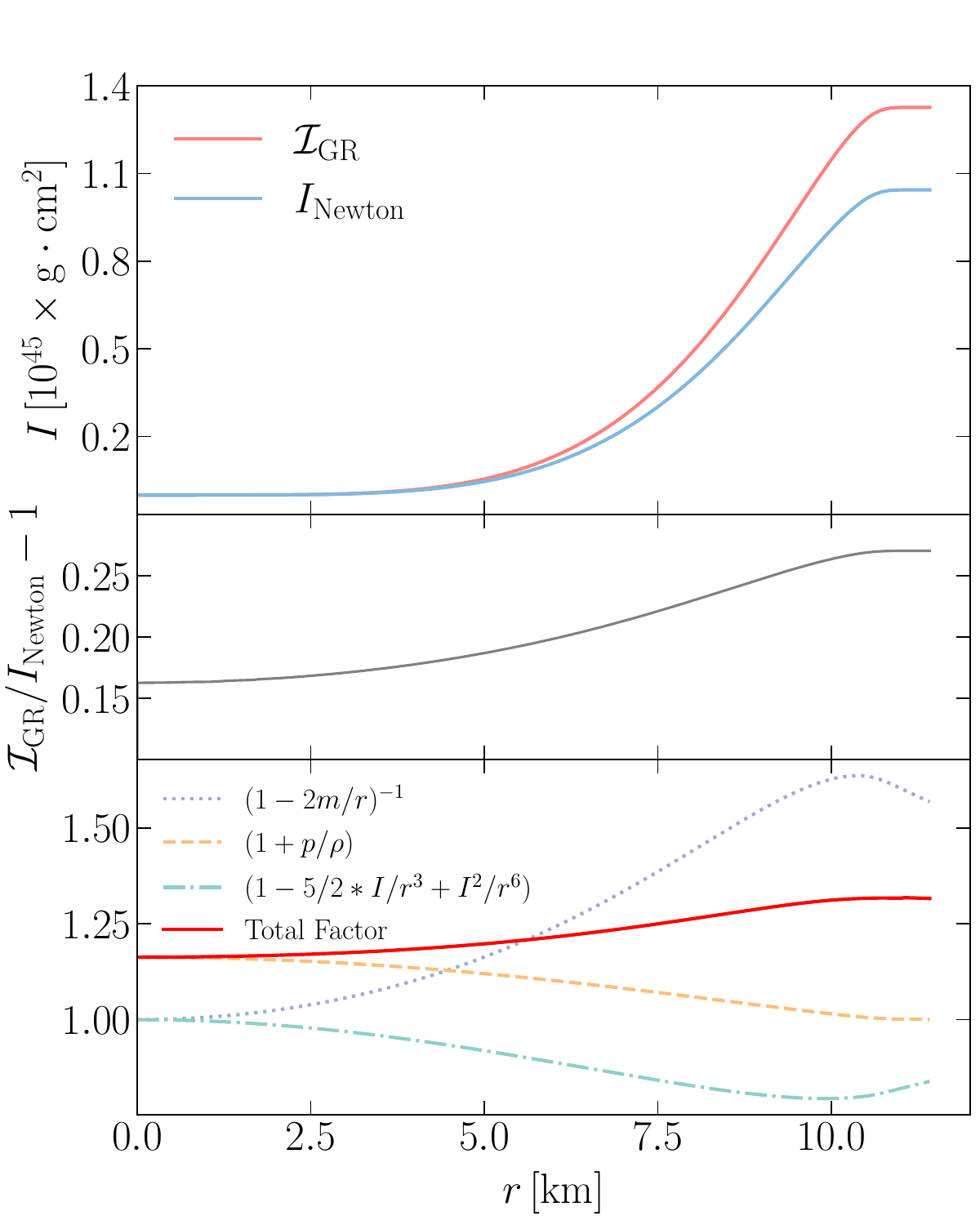}
    \caption{The MOI function of a $1.4\,{\rm M_\odot}$ NS with EOS AP4. The top
    panel represents the MOI function in GR, $\mathcal{I}_{\rm GR}$, and the MOI
    in Newtonian gravity, $I_{\rm Newton}$. The middle panel represents the
    relative difference between $\mathcal{I}_{\rm GR}$ and $I_{\rm Newton}$. The
    bottom panel represents the magnitude of the dimensionless correction
    factors in Eq.~(\ref{Eq_dIdr}). The ``Total Factor'' is the product of three
    correction factors.}
    \label{Fig_1}
\end{figure}

We consider a NS of $1.4\,{M_\odot}$ with EOS  AP4 as an example, and calculate
the MOI function $\mathcal{I}$ with respect to $r$. The result is shown in Fig
\ref{Fig_1}. Besides, we calculate the MOI in the framework of Newtonian gravity
according to
\begin{equation}
    \frac{\mathrm{d} I_{\rm Newton}}{\mathrm{d} r}=\frac{8}{3} \pi r^4 \rho\,.
\label{Eq_dIdr_Newton}
\end{equation}
The difference in the MOI between GR and the Newtonian gravity reaches
$\sim20\%$, which is consistent with the value of the dimensionless factors.
Furthermore, the calculation of MOI in GR is no longer linear, because in the
right-hand side of Eq.~(\ref{Eq_dIdr}), there exists a factor related to
$\mathcal{I}$ breaking the linearity. That is to say, if the MOI of a sphere is
$I_{1}$ and the MOI of another concentric sphere is $I_{2}$, then the total MOI
is not simply $I_{1}+I_{2}$ as in the case of Newtonian gravity.

\section{MOI of axisymmetric NSs in the SME}
\label{Sec3_MOI_SME}

In this section, we derive the modified PDE in the SME to calculate the MOI for
axisymmetric NSs. In the axisymmetric case, the PDE becomes more complicated in
its dependence on $\theta$, making it difficult to solve the PDE through
separation of variables. We analyze the asymptotic behavior of the solution and
solve it numerically with the finite element method.  

\subsection{NSs in the SME}
\label{Sec31_NS_SME}

In the gravitational sector of the minimal SME, the linearized field equations
can be written as~\cite{Bailey:2006fd, Bailey:2014bta}
\begin{equation}
    G_{\mu \nu}=8 \pi T_{\mu \nu}-\bar{s}^{\alpha \beta} G_{\mu \alpha \beta \nu}\,,
\label{Eq_Field_Eq_SME}
\end{equation}
where
\begin{equation}
\begin{aligned}
    G_{\alpha \beta \gamma \delta}=&-R_{\alpha \beta \gamma \delta}+g_{\alpha
    \gamma} R_{\beta \delta}+g_{\beta \delta} R_{\alpha \gamma}-g_{\alpha
    \delta} R_{\beta \gamma}\\ 
    & -g_{\beta \gamma} R_{\alpha \delta}-\frac{1}{2}\left(g_{\alpha \gamma}
    g_{\beta \delta}-g_{\alpha \delta} g_{\beta \gamma}\right) R\,,
\end{aligned}
\end{equation}
and $\bar{s}^{\alpha \beta}$ are the Lorentz violation coefficients. By applying
it to the strong-field regime, we acknowledge that higher-order corrections at
${\cal O}( s^2 \cdot h)$ and ${\cal O}(s\cdot h^2)$ might exist, where $s$ is
the typical value of $\bar{s}^{\alpha \beta}$ components and $h$ is the typical
metric deviation from the flat spacetime. For NSs, $h\sim 0.1$ while limits on
$\bar{s}^{\alpha \beta}$ are as low as $10^{-11}$~\cite{Shao:2014oha,
Kostelecky:2008ts}. If we briefly assume $s \lesssim h \sim 0.1$, then the
relative error caused by applying Eq.~(\ref{Eq_Field_Eq_SME}) to NSs is less
than ${\cal O}(s\cdot h) \lesssim 1\%$. As a starting study for testing Lorentz
symmetry using future precise measurements of NSs' MOI, we can tolerate the
error for now. Our formalism to be presented can be directly applied to a more
general version of Eq.~(\ref{Eq_Field_Eq_SME}) where higher-order terms are
included.

Using Eq.~(\ref{Eq_Field_Eq_SME}), the Lorentz-violating term $\bar{s}^{\alpha
\beta} G_{\mu \alpha \beta \nu}$ results in corrections to the metric of the NS.
The modified metric can be represented as
\begin{equation}
    g_{\mu\nu} = g_{\mu\nu}^{\rm{\rm{LI}}} + \delta g_{\mu\nu}^{\rm LV}\,,
\end{equation}
where 
\begin{equation}
    g_{\mu\nu}^{\rm{LI}} = {\rm diag}\Big\{-e^{\nu}, \, e^{\lambda}, \, r^2, \,
    r^2\sin{\theta}\Big\}\,
\end{equation}
is the TOV solution for the NS in GR, and 
\begin{equation}
    \delta g_{\mu\nu}^{\rm{LV}} = {\rm diag} \Big\{-\delta\phi (r,\theta), \, 0,
    \, 0, \, 0 \Big\}\,
\end{equation}
represents the correction caused by Lorentz violation.
We have~\cite{Bailey:2006fd},
\begin{equation}
    \delta\phi (r,\theta) =-\bar{s}^{jk}U^{jk}\,,
    \label{lvcorreq}
\end{equation}
and
\begin{equation}
    U^{jk} = \int \frac{ \big(x^j - x^{\prime j} \big) \big(x^k - x^{\prime
    k}\big)}{\left| \vv{x} - \vv{x}^{\prime}\right|^{3}} \rho
    \big(\vv{x}^{\prime} \big) \mathrm{d}^{3} x^{\prime}\,,
\end{equation}
where $\rho$ is the energy density distribution in the NS. Note that the
repeated indices in Eq.~(\ref{lvcorreq}) are summed over.

Under the perturbation of Lorentz violation, the structure of a NS is
affected~\cite{Xu:2019gua}. The energy density and the pressure are given by
\begin{align}
    \rho &= \rho^{(0)} + \rho^{(1)}\,, \\
    p &= p^{(0)} + p^{(1)}\,,
\end{align}
where $\rho^{(0)}(r)$ and $p^{(0)}(r)$ are respectively the energy density and
pressure of undisturbed NSs, and $\rho^{(1)}(\vv{x})$ and $p^{(1)}(\vv{x})$ are
corresponding corrections caused by Lorentz violation, which
are~\cite{Xu:2019gua} 
\begin{align}
    \rho^{(1)}(\vv{x}) &=-\alpha(\theta, \varphi) r \partial_r \rho^{(0)}(r)\,, \\
    p^{(1)}(\vv{x}) &=-\alpha(\theta, \varphi) r \partial_r p^{(0)}(r)\,.
\end{align}
In the above equations, $\alpha(\theta, \varphi)$ is,
\begin{equation}
    \alpha(\theta, \varphi)=\frac{1}{2} \sum_{m=-2}^2 s_{2 m}^{(s)} Y_{2
    m}(\theta, \varphi)\,,
\end{equation}
where $Y_{lm}(\theta,\varphi)$ are the spherical harmonics, and $s_{2 m}^{(s)}$
are
\begin{equation}
\begin{aligned}
& s_{2,-2}^{(s)}=\sqrt{\frac{2 \pi}{15}}\left(\bar{s}^{x x}-\bar{s}^{y y}+2 i \bar{s}^{x y}\right)\,, \\
& s_{2,-1}^{(s)}=2 \sqrt{\frac{2 \pi}{15}}\left(\bar{s}^{x z}+i \bar{s}^{y z}\right)\,, \\
& s_{2,0}^{(s)}=\frac{2}{3} \sqrt{\frac{\pi}{5}}\left(-\bar{s}^{x x}-\bar{s}^{y y}+2 \bar{s}^{z z}\right)\,, \\
& s_{2,1}^{(s)}=2 \sqrt{\frac{2 \pi}{15}}\left(-\bar{s}^{x z}+i \bar{s}^{y z}\right)\,, \\
& s_{2,2}^{(s)}=\sqrt{\frac{2 \pi}{15}}\left(\bar{s}^{x x}-\bar{s}^{y y}-2 i \bar{s}^{x y}\right)\,.
\end{aligned}
\end{equation}

We study the specific case with axial symmetry, so we choose  $\bar{s}^{\mu\nu}$
to vanish except for $\bar{s}^{zz}$ where $z$-axis is the NS's spin direction.
In order words, we assume that the specific direction of the Lorentz-violating
field is parallel to the spinning axis of the NS. 

In a brief summary, modifications from Lorentz violation are categorized into
two aspects. Firstly, Lorentz violation modifies the gravitational field
equations. Secondly, it induces deformations in the NS structure.

\subsection{The modified PDE for MOI}
\label{Sec32_Mod_PDE}

Similar to the calculation in GR, the perturbative rotational properties in a
stationary axisymmetric background spacetime are governed by the $t\varphi$
component of Eq.~(\ref{Eq_Field_Eq_SME}). We can begin with the general form of
an axisymmetric metric as given in Eq.~(\ref{Eq_metric_Ax}). We adopt the
slow-rotation assumption, so that at the linear order of the angular velocity
$\Omega$, the functions for the diagonal metric components take the background
result while the function $L$ takes the form of Eq.~(\ref{Eq_L_exp}), and the
related function $\bar{\omega}$ can be defined as in
Eq.~(\ref{Eq_defination_baromega}). Furthermore, we now would like to consider
the leading corrections due to the Lorentz violation coefficient $\bar{s}^{zz}$,
so we can treat $\bar{\omega}$ as having an expansion in terms of
$\bar{s}^{zz}$,
\begin{equation}
    \bar{\omega} = \bar{\omega}^{(0)} + \bar{\omega}^{(1)}+ {\cal O} \big(|\bar
    s^{zz}|^2 \big)\,,
\end{equation}
where $\bar{\omega}^{(0)}$ corresponds to the GR result with vanishing
$\bar{s}^{zz}$, and $\bar{\omega}^{(1)}$ gives the leading-order correction to
the metric component $g_{t\varphi}$ due to Lorentz violation.

To calculate $\bar{\omega}^{(1)}$, we write out the $t\varphi$ component of
Eq.~(\ref{Eq_Field_Eq_SME}) and arrange it in orders of $\bar{s}^{zz}$. The
zeroth-order equation in $\bar{s}^{zz}$ reads
\begin{widetext}
\begin{equation}
    \frac{1}{r^4}\frac{\partial}{\partial r}\left(r^4 j \frac{\partial
    \bar{\omega}^{(0)}}{\partial r} \right) + \frac{4}{r} \frac{\mathrm{d}
    j}{\mathrm{d} r} \bar{\omega}^{(0)} +
    \frac{e^{(\lambda-\nu)/2}}{r^2}\frac{1}{\sin^{3}{\theta}}
    \frac{\partial}{\partial \theta} \left( \sin^{3}{\theta} \frac{\partial
    \bar{\omega}^{(0)}}{\partial \theta} \right)=0\,,
\label{Eq_MOI_ZeroOrder}
\end{equation}
which is identical to Eq.~(\ref{Eq_GR_MOI}) as expected, and therefore
$\bar{\omega}^{(0)}$ is just the GR solution.  The first-order equation in
$\bar{s}^{zz}$ is
\begin{equation}
    \frac{1}{r^4} \frac{\partial}{\partial r}\left(r^4 j \frac{\partial
    \bar{\omega}^{(1)}}{\partial r} \right) + \frac{4}{r} \frac{\mathrm{d}
    j}{\mathrm{d} r} \bar{\omega}^{(1)} +
    \frac{e^{(\lambda-\nu)/2}}{r^2}\frac{1}{\sin^{3}{\theta}}
    \frac{\partial}{\partial \theta} \left( \sin^{3}{\theta} \frac{\partial
    \bar{\omega}^{(1)}}{\partial \theta} \right) 
     = S_{1}(r,\theta)+S_{2}(r,\theta)\,,
\label{Eq_MOI_PDE}
\end{equation}
where
\begin{align}
    S_1(r,\theta) &= \frac{4}{r}\frac{\mathrm{d} j}{\mathrm{d} r}
    \bar{\omega}^{(0)} \delta\phi e^{-\nu} + 16 \pi \Big(\rho^{(1)}+p^{(1)}
    \Big) e^{(\lambda-\nu)/2} \bar{\omega}^{(0)}\,, \label{Eq_S1} \\
     S_2(r,\theta) &= \left(\delta\phi e^{-\nu} +
     \bar{s}^{zz}\sin^{2}{\theta}\right) \frac{1}{r^4}\frac{\partial}{\partial
     r}\left(r^4 j \frac{\partial \bar{\omega}^{(0)}}{\partial r}\right) +
     \frac{1}{2} j \left[\delta\phi_{,r}e^{-\nu} - \delta\phi e^{-\nu}
     \frac{\mathrm{d} \nu}{\mathrm{d} r} + \frac{\bar{s}^{zz}}{r}
     \left(8e^{\lambda}\cos^{2}{\theta}-2\sin^{2}{\theta}\right)\right]\frac{\partial
     \bar{\omega}^{(0)}}{\partial r}\,.  \label{Eq_S2}
\end{align}
Note that $S_1(r,\theta)$ arises from modifications in $T_{t\varphi}$ due to
Lorentz violation while $S_2(r,\theta)$ arises from $G_{t\varphi}$ and the
Lorentz-violating term $\bar s^{zz} G_{tzz\varphi}$ in the field equation
(\ref{Eq_Field_Eq_SME}).  With the boundary conditions in Eq.~(\ref{Eq_Bound1})
and Eq.~(\ref{Eq_Bound2}) for
$\bar{\omega}\approx\bar{\omega}^{(0)}+\bar{\omega}^{(1)}$, $\bar{\omega}^{(1)}$
can be solved from Eq.~(\ref{Eq_MOI_PDE}) once we have the GR solution
$\bar{\omega}^{(0)}$.

Numerical method needs to be employed to solve Eq.~(\ref{Eq_MOI_PDE}) as well as
$\bar{\omega}^{(0)}$ inside the NS. Instead of solving $\bar{\omega}^{(0)}$ and
$\bar{\omega}^{(1)}$ one by one, it is straightforward to solve the combined PDE
\begin{equation}
    \frac{1}{r^4}\frac{\partial}{\partial r}\left(r^4 j \frac{\partial
    \bar{\omega}}{\partial r} \right) + \frac{4}{r} \frac{\mathrm{d}
    j}{\mathrm{d} r} \bar{\omega}  +
    \frac{e^{(\lambda-\nu)/2}}{r^2}\frac{1}{\sin^{3}{\theta}}
    \frac{\partial}{\partial \theta} \left( \sin^{3}{\theta} \frac{\partial
    \bar{\omega}}{\partial \theta} \right)  =
    S^{\prime}_{1}(r,\theta)+S^{\prime}_{2}(r,\theta)\,, \label{Eq_PDE_Comb2}
\end{equation}
where
\begin{align}
     S^{\prime}_1(r,\theta) &= \frac{4}{r}\frac{\mathrm{d} j}{\mathrm{d} r}
     \bar{\omega} \delta\phi e^{-\nu} + 16 \pi \Big(\rho^{(1)}+p^{(1)} \Big)
     e^{(\lambda-\nu)/2} \bar{\omega}\,, \\
     S^{\prime}_2(r,\theta)  &= \left(\delta\phi e^{-\nu} +
     \bar{s}^{zz}\sin^{2}{\theta}\right) \frac{1}{r^4}\frac{\partial}{\partial
     r}\left(r^4 j \frac{\partial \bar{\omega}}{\partial r}\right) + \frac{1}{2}
     j \left[\delta\phi_{,r}e^{-\nu} - \delta\phi e^{-\nu} \frac{\mathrm{d}
     \nu}{\mathrm{d} r} + \frac{\bar{s}^{zz}}{r}
     \left(8e^{\lambda}\cos^{2}{\theta}-2\sin^{2}{\theta}\right)\right]\frac{\partial
     \bar{\omega}}{\partial r}\,.
\end{align}
\end{widetext}

Before we proceed to the numerical results, let us clarify the asymptotic
behavior of $\bar{\omega}$ so that the angular momentum and hence the MOI can be
defined.  

We split the solution of Eq.~(\ref{Eq_PDE_Comb2}) into two parts, $\bar{\omega}
= \bar{\omega}_1 + \bar{\omega}_2$, where $\bar{\omega}_1$ is the solution
ignoring the source $S'_2$ and it contains the zeroth-order solution in
$\bar{s}^{zz}$, which represents the contribution from the matter, and
$\bar{\omega}_2$ is the solution ignoring the source $S'_1$ and it does not
contain the zeroth-order solution in $\bar{s}^{zz}$, which represents the
contribution from the modification of the gravity theory. We discuss the
asymptotic behaviors of $\bar{\omega}_1$ and $\bar{\omega}_2$ separately.


Outside the star, $S'_1$ is zero, and hence $\bar{\omega}_1$ has the exterior
solution~\cite{Hartle:1967he}
\begin{equation}
\bar{\omega}_1 = \sum\limits_{l=1}^{\infty} \bar{\omega}_{1l} \left(  -
\frac{1}{\sin\theta} \frac{dP_l}{d\theta} \right) ,
\label{omega2extsol}
\end{equation}
where $P_l$ are the Legendre polynomials and $\bar{\omega}_{1l}$ consists of the
$r^{-l-2}$ terms and the $r^{l-1}$ terms. Because of the asymptotic flatness
boundary condition, the $r^{l-1}$ terms with $l \ne 1$ must have vanishing
coefficients. Now very different from GR, the PDE for $\bar{\omega}_1$ does not
admit separation of variables inside the NSs due to the source term $S'_1$, so
for each $l$ in $\bar{\omega}_{1l}$, the coefficient for $r^{-l-2}$ is no longer
forced to be proportional to the coefficient for $r^{l-1}$, meaning that the
$r^{-l-2}$ terms can exist in spite of the absence of the $r^{l-1}$ terms. 

As axisymmetric NSs exhibit reflection symmetry about the equatorial plane, we
have $\bar{\omega}(r,\theta)=\bar{\omega}(r,\pi-\theta)$, which implies that
odd-power terms of $\cos{\theta}$ should not appear in $\bar{\omega}$ and hence
$\bar\omega_1$, and it excludes the even $l$ terms in Eq.~(\ref{omega2extsol}).
In conclusion, the expansion of $\bar\omega_1$ outside the star takes the form
\begin{equation}
\bar{\omega}_1 = {\cal{A}}_1 + \frac{{\cal{B}}_1}{r^3} + \frac{{\cal{C}}_1}{r^5}
\left(  -  \frac{1}{\sin\theta} \frac{dP_3}{d\theta} \right) + O\left(
\frac{1}{r^7} \right),
\label{omega2asymp}
\end{equation}
where ${\cal{A}}_1, \, {\cal{B}}_1, \, {\cal{C}}_1$ are constants.

For $\bar{\omega}_2$, its PDE neither admits separation of variables outside the
star nor inside the star. Then we have to assume its asymptotic expansion in
terms of $1/r$ to be the general form, 
\begin{equation}
    \bar{\omega}_2 =
    {\mathcal{A}}_2+\frac{{\mathcal{B}}_2(\theta)}{r^3}+\frac{{\mathcal{C}}_2(\theta)}{r^4}+\dots
    \,, \label{omega1asymp}
\end{equation}
Note that there are no $1/r$ term nor $1/r^{2}$ term because of the boundary
condition of asymptotic flatness. For the very same boundary condition, the
coefficient ${\mathcal{A}}_2$ has to be independent of $\theta$ and satisfies
$\Omega = {\mathcal{A}}_1 + {\mathcal{A}}_2$, where ${\mathcal{A}}_1$ is the
constant in Eq.~(\ref{omega2asymp}). 

Combining the asymptotic expansions in
Eqs.~(\ref{omega2asymp}--\ref{omega1asymp}), substituting into
Eq.~(\ref{Eq_PDE_Comb2}), and arranging the equation in orders of $1/r$, we find
at the leading order an equation for the coefficient ${\mathcal{B}}_2(\theta)$,
\begin{equation}
    \frac{1}{\sin^{3}{\theta}}\frac{\mathrm{d}}{\mathrm{d}
    \theta}\left(\sin^{3}\theta \frac{\mathrm{d} \mathcal{B}_2}{\mathrm{d}
    \theta} \right) = 3 \bar s^{zz} {\mathcal{B}}_1 \big(\sin^{2}\theta - 4
    \cos^{2}\theta \big)\,,
\end{equation}
and equations for the coefficient ${\mathcal{C}}_2(\theta)$ are at higher
orders.  Note that we have used $\bar{\omega}_2/\bar{\omega}_1 \sim O\left(\bar
s^{zz}\right)$ and substituted $\bar{\omega}$ with $\bar{\omega}_1$ in $S'_2$.
For $\bar{\omega}_2$ to be smooth, we impose 
\begin{equation}
    \left.\frac{{\rm d} {\mathcal{B}_2}(\theta)}{{\rm d} \theta} \right|_{\theta
    = 0} = \left.\frac{{\rm d} {\mathcal{B}_2}(\theta)}{{\rm d} \theta}
    \right|_{\theta = \pi} = 0\, ,
\label{Eq_Bound3}
\end{equation}
and get 
\begin{equation}
    {\mathcal{B}_2}(\theta) = -\frac{3}{2} \bar s^{zz} {\mathcal{B}}_1
    \sin^{2}\theta + \mathrm{constant} \, .  \label{Eq_B_solution}
\end{equation}
Therefore, gathering Eqs.~(\ref{omega2asymp}--\ref{omega1asymp}) we find the
asymptotic behavior of $\bar{\omega}$ to be
\begin{equation}
    \bar{\omega} = \Omega \left[ 1 - \frac{2}{r^{3}}\left(p \sin^{2}\theta + q
    \right) + \mathcal{O}\left(\frac{1}{r^4}\right) \right], \label{omegaasymp}
\end{equation}
where $\Omega, \, p$, and $q$ are constants.  

With the asymptotic expression for $\bar\omega$ in Eq.~(\ref{omegaasymp}), the
angular momentum of the spacetime is found to be~\cite{Misner:1973prb},
\begin{equation}
    J = \Omega\left(\frac{4}{5} p + q\right)\,.
\end{equation}
Then the MOI of the star is
\begin{equation}
I = \frac{J}{\Omega} = \left(\frac{4}{5} p + q\right) .
\label{moieq}
\end{equation} 
Now we are ready to calculate the constants $p, \, q$ and therefore the MOI for
NSs  by numerically solving Eq.~(\ref{Eq_PDE_Comb2}). 

\subsection{Numerical calculation}
\label{Sec33_Num_Cal}

To numerically solve the PDE for $\bar{\omega}$, we first perform a change of
variables which is inspired by \citet{Cook:1993qr},
\begin{equation}
    (r, \, \theta) \rightarrow \left(x \equiv \frac{r\cos{\theta}}{r+R}, \,
    y\equiv \frac{r\sin{\theta}}{r+R} \right)\,, \label{Eq_Change_Var}
\end{equation}
where $R$ is the radius of an unperturbed NS.  Equation~(\ref{Eq_PDE_Comb2})
then changes to
\begin{widetext}
\begin{equation}
\begin{aligned}
   f_{1}(x,y) \left(x\bar{\omega}_{,x} + y\bar{\omega}_{,y}\right) & +
   f_{2}(x,y) \left( x^2 \bar{\omega}_{,xx}+2xy \bar{\omega}_{,xy} + y^2
   \bar{\omega}_{,yy}\right)  +  f_{3}(x,y) \bar{\omega} \\ 
   & + f_{4}(x,y) \left( x\bar{\omega}_{,y} - y \bar{\omega}_{,x} \right) +
   f_{5}(x,y) \left( y^2 \bar{\omega}_{,xx} -2xy \bar{\omega}_{,xy} + x^2
   \bar{\omega}_{,yy} - x \bar{\omega}_{,x} - y \bar{\omega}_{,y} \right)  =
   0\,,
\end{aligned}
\label{Eq_MOI_GR_xy}
\end{equation}
where
\begin{align}
    f_{1}(x,y) = & \left[ 4j(1-\bar{r})^{3} + \bar{r}(1-\bar{r})^{2}R
    \frac{\mathrm{d} j }{\mathrm{d} r} - 2j\bar{r}(1-\bar{r})^{3}\right]\left(
    \bar{r}^{2} - \bar{r}^{2} \delta\phi e^{-\nu} - \bar{s}^{zz} y^{2} \right)
    \nonumber \\
    & - \left[ \frac{1}{2} \delta \phi_{,r} e^{-\nu} j R \bar{r}^{3}
    (1-\bar{r})^{2} - \frac{1}{2} \delta \phi e^{-\nu} j R \frac{\mathrm{d}
    \nu}{\mathrm{d} r} \bar{r}^{3} (1-\bar{r})^{2} - \frac{1}{2} \bar{s}^{zz} j
    (1-\bar{r})^{3}(-8 e^{\lambda} x^{2} + 2 y^{2})\right]\,, \label{Eq_f1} \\
    f_{2}(x,y) =& j (1-\bar{r})^{4} \left( \bar{r}^{2} - \bar{r}^{2} \delta\phi
    e^{-\nu} - \bar{s}^{zz} y^{2} \right)\,, \label{Eq_f2} \\
    f_{3}(x,y) =& 4 \bar{r}^{3} (1-\bar{r}) R \frac{\mathrm{d} j}{\mathrm{d} r}
    (1-\delta\phi e^{-\nu}) - 16 \pi \left( \rho^{(1)} + p^{(1)} \right)
    e^{(\lambda-\nu)/2} \bar{r}^{4} R^{2}\,, \\
    f_{4}(x,y) =& e^{(\lambda-\nu)/2} \bar{r}^{2} (1-\bar{r})^{2}
    \frac{3x}{y}\,, \\
    f_{5}(x,y) =& e^{(\lambda-\nu)/2} \bar{r}^{2} (1-\bar{r})^{2}\,,
    \label{Eq_f5}
\end{align}
\end{widetext}
with $\bar{r} = r/(r+R) = \sqrt{x^{2}+y^{2}}$ being a dimensionless variable.
Axisymmetric NSs  exhibit reflection symmetry about the equatorial plane, so the
parameter space of $(x,y)$ can be reduced to a quarter-sector, satisfying $x>=0$
and $y>=0$. The boundary conditions become
\begin{align}
    \left.\frac{\partial \bar{\omega}}{\partial x}\right|_{x=0} &=
    \left.\frac{\partial \bar{\omega}}{\partial y}\right|_{y=0} = 0\,,
    \label{Eq_Bound5} \\
    \bar{\omega}\big|_{\bar{r}\rightarrow 1} &= \Omega\,.
    \label{Eq_Bound6}
\end{align}

The advantage of the $(x,y)$ variables are in two aspects. First, $r \in
\left(0,+\infty \right)$ corresponds to $\bar{r} \in \left(0,1\right)$. We can
use a finite sector of unit radius on the $xy$-plane to represent the infinite
$(r,\theta)$ plane, which is convenient for us to input boundary conditions at
infinity in numerics. Additionally, $r=R$ corresponds to $\bar{r}=1/2$. As we
use finite element method with uniform grids in the $xy$-plane, the interior of
the NS will be solved more meticulously compared to using the $(r,\theta)$
variables. This is beneficial for increasing the accuracy. Second, we choose
$(x,y)$ instead of $(\bar{r},\theta)$ to avoid inputting the regular condition
similar to Eq.~(\ref{Eq_Bound1}) when $\bar{r}$ tends to zero.

Using the variables $(x, y)$, we find that finite element method suffices to
solve the PDE for $\bar \omega$ in Eq.~(\ref{Eq_MOI_GR_xy}). After each
numerical solution is obtained, we fit it at large $r$ according to
Eq.~(\ref{omegaasymp}) to extract the constants: $\Omega, \, p$, and $q$.
Afterwards the MOI for the star is calculated using Eq.~(\ref{moieq}).

\begin{figure}
    \centering
    \includegraphics[scale=0.43, trim=30 30 0 65]{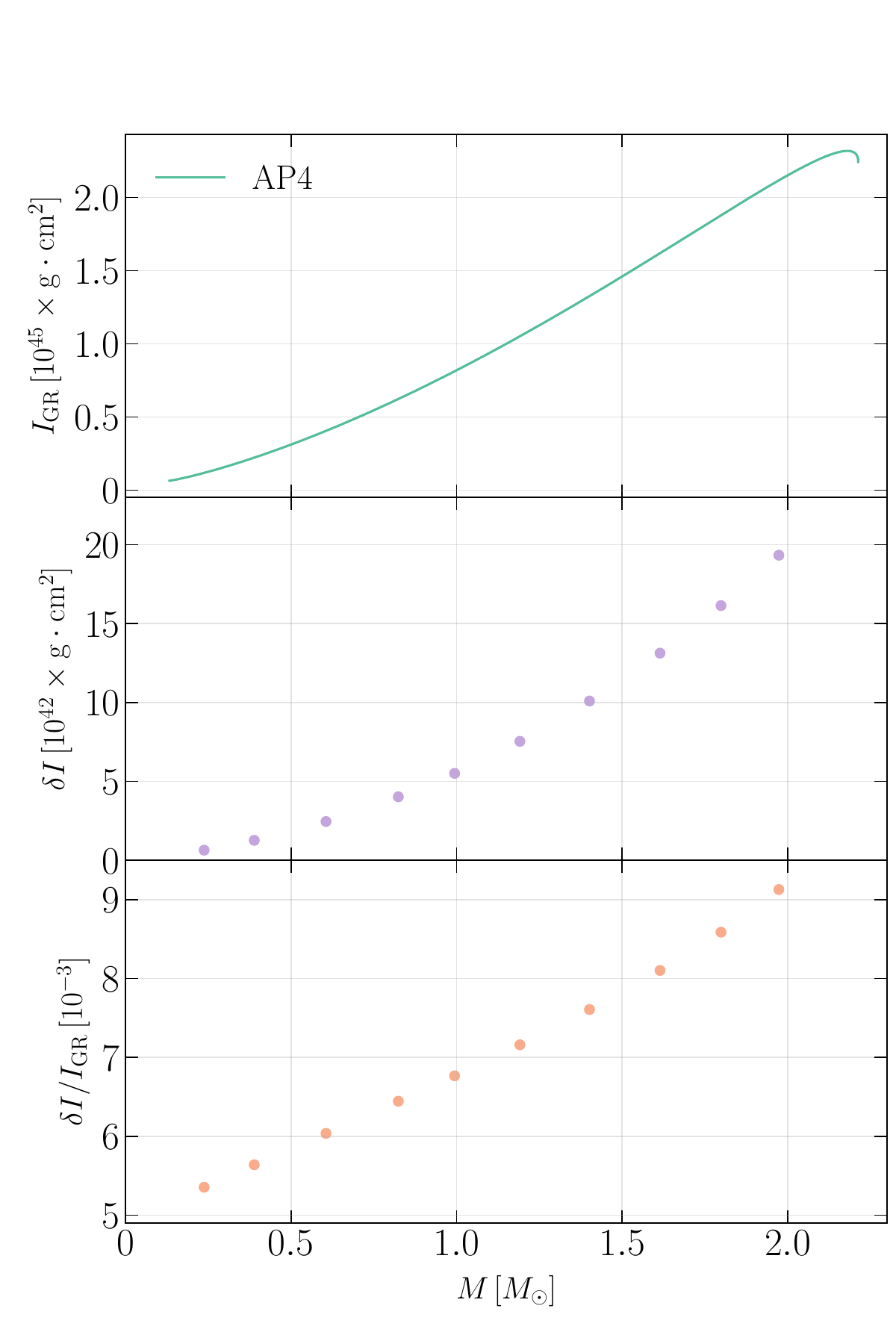}
    \caption{Corrections to the MOIs of NSs caused by Lorentz violation with EOS
    AP4, as functions of the mass of the NS. The top panel shows the MOI  of
    spherical NSs in GR, $I_{\rm GR}$. The middle panel shows the absolute
    correction to the MOI caused by Lorentz violation with $\bar{s}^{zz} =
    10^{-2}$. The bottom panel shows the ratio $\delta I/I_{\rm GR}$.}
    \label{Fig_2}
\end{figure}

We use the EOS AP4 as an example and calculate MOIs for NSs with different
masses. The results are shown in Fig.~\ref{Fig_2}. For the illustration purpose,
we have taken $\bar s^{zz} = 10^{-2}$ in our numerical calculation. From Fig.~\ref{Fig_2}, we can see that the ratio $\delta I/I_{\rm
GR}$ follows a relatively good linear relation. It may help us to quickly
estimate $\delta I$.

We have completed the calculation of MOIs for axisymmetric NSs due to Lorentz
violation. We find that it is also interesting to calculate the MOI solely from
$\bar\omega_1$. By doing this we ignore the source term $S'_2$ in
Eq.~(\ref{Eq_PDE_Comb2}), so the Lorentz-violating effect comes into play only
through the energy-momentum tensor of the NS matter. The result can be compared
with the estimation made by \citet{Xu:2019gua}, where the correction in the MOI
caused by Lorentz violation is calculated in the Newtonian way by only
considering the change in matter distribution, via
\begin{equation}
    \delta I_{\rm Newton} = \int \rho^{(1)}(r,\theta)r^{2}\sin^{2}\theta\,
    \mathrm{d}^3x = -\frac{1}{3} \bar s^{zz} I_{\rm Newton},
\end{equation}
where $I_{\rm Newton}$ is the Newtonian MOI with the mass density $\rho^{(0)}$
in the absence of Lorentz violation.

\begin{figure}
    \centering
    \includegraphics[scale=0.43, trim=5 5 0 0]{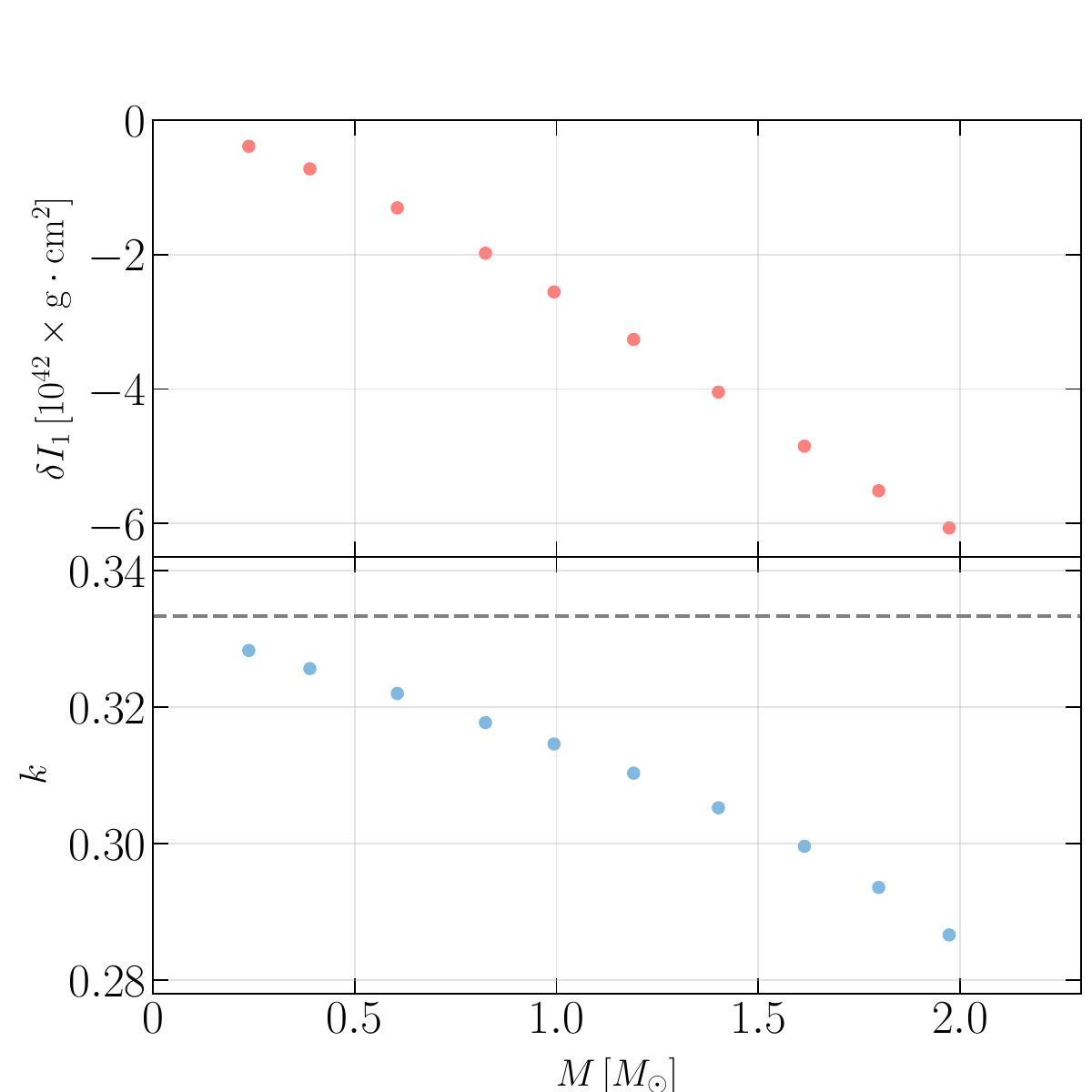}
    \caption{Corrections to MOIs caused by deformations of NSs with EOS AP4, as
    functions of the NS mass. The top panel shows the correction to MOI caused
    by deformations of NSs with $\bar{s}^{zz} = 10^{-2}$. The bottom panel shows
    the value of $k$ defined in Eq.~(\ref{Eq_Def_k}). The grey dashed line
    represents $k=1/3$, which is the Newtonian limit for $k$.}
    \label{Fig_3}
\end{figure}

The numerical approach to calculate $\bar\omega_1$ differs slightly from what we
have done for calculating $\bar\omega$. We only need to remove the terms
corresponding to $S'_2$ in Eqs.~(\ref{Eq_MOI_GR_xy}--\ref{Eq_f5}). After
obtaining numerical solutions and fitting them according to
Eq.~(\ref{omega2asymp}), the MOI is then calculated by  $I_1 =
-{{\cal{B}}_1}/{2{\cal{A}}_1}$.  For NSs with different masses, results of $I_1$
are shown in Fig.~\ref{Fig_3} in terms of $\delta I_1 = I_1 - I_{\rm GR}$.
Figure~\ref{Fig_3} also shows the change of a factor $k$ defined in
\begin{equation}
\delta I_1 = -k \bar s^{zz} I_{\rm GR} .
\label{Eq_Def_k}
\end{equation} 
As the mass of the NS decreases, we expect $I_{\rm GR} \rightarrow I_{\rm
Newton}$ and $\delta I_1 \rightarrow \delta I_{\rm Newton}$ so that
$k\rightarrow 1/3$. This is exactly what we see in Fig.~\ref{Fig_3}. 

\section{Summary}
\label{Sec5_Summary}

In this paper, we develop the method to calculate axisymmetric NSs' MOI in the
presence of Lorentz violation in a relativistic setting.  Solutions are worked
out for the first time in the effective-field-theoretic framework of SME.  We
treat the effect of Lorentz violation as a perturbation and derive the modified
PDE for the MOI from the gravitational field equations. Then, we discuss the
asymptotic behavior of the solution analytically. After that, we perform a
change of variables, solve the PDE with the finite element method and fit the
numerical solutions with polynomials to get the MOI.

After obtaining the numerical results, we calculate correction to the MOI of
NSs caused by Lorentz violation. Besides that, we separately calculate
correction to the MOI caused by the deformation of the NS. We compare the ratio
$\delta I_{1}/I_{\rm GR}$ with its counterpart in Newtonian gravity and show the
difference. For a 1.4 $M_{\odot}$ NS with EOS AP4, the difference is at the
level of $\sim 8\%$.

In the future, we can extend this method to study the structure of axisymmetric
NSs in other modified gravity theories, e.g.\ the bumblebee
theory~\cite{Bailey:2006fd, Xu:2022frb} and the Einstein-\AE{}ther
theory~\cite{Jacobson:2000xp}. \Reply{If stable axisymmetric NS solutions exist in these modified gravity theories, the calculation procedure outlined in this paper may serve as an important reference for calculating the MOI of axisymmetric NSs in these modified gravity theories. This assists us in studying the properties of NSs in the modified gravity theories. As a byproduct, we have also presented a consistent method for calculating the MOI corrections caused by axisymmetric deformations in NSs within the framework of GR, which may offer a more precise method for computing MOI corrections in some theoretical models as well. } We are looking forward to calculating
corrections to the MOI caused by general deformations of NSs in the framework of
GR, to help us understand the physics of NSs better.

\begin{acknowledgments}
We thank Yong Gao and Hong-Bo Li for useful discussions.  This work was
supported by the National SKA Program of China (2020SKA0120300), the National
Natural Science Foundation of China (11975027, 11991053),  the Max Planck
Partner Group Program funded by the Max Planck Society, and the High-Performance
Computing Platform of Peking University. Y.D.\ was supported by the National
Training Program of Innovation for Undergraduates at Peking University.
\end{acknowledgments}

\appendix

\bibliography{MOI_SME}

\end{document}